\documentclass[aps,twocolumn,showpacs,byrevtex]{revtex4}
\usepackage{times}
\usepackage{amsmath}
\usepackage{graphicx}
\usepackage{color}
\newcommand{\y}{Y(4260)}
\newcommand{\z}{Z_c(3900)}
\newcommand{\x}{X(3872)}
\newcommand{\pp}{\pi^+\pi^-}

\newcommand{\LL}{\ell^+\ell^-}
\newcommand{\EE}{e^+e^-}
\newcommand{\MM}{\mu^+\mu^-}

\newcommand{\etap}{\eta^\prime}
\newcommand{\psip}{\psi(3686)}

\newcommand{\jpsi}{J/\psi}

\newcommand{\ppjpsi}{\pi^+\pi^-J/\psi}

\def\Journal#1#2#3#4{{#1} {\bf #2}, #3 (#4)}

\def\PLB{Phys. Lett. B}
\def\PRL{Phys. Rev. Lett.}
\def\PRD{Phys. Rev. D}

\def\EPJC{Eur. Phys. J. C}

\begin{document}


\title{
\boldmath Observation of $\EE\to \gamma\x$ at BESIII}

\author{
M.~Ablikim$^{1}$, M.~N.~Achasov$^{8,a}$, X.~C.~Ai$^{1}$, O.~Albayrak$^{4}$, D.~J.~Ambrose$^{41}$, F.~F.~An$^{1}$, Q.~An$^{42}$, J.~Z.~Bai$^{1}$, R.~Baldini Ferroli$^{19A}$, Y.~Ban$^{28}$
% ...
\\
\vspace{0.2cm}
(BESIII Collaboration)\\
\vspace{0.2cm} {\it
$^{1}$ Institute of High Energy Physics, Beijing 100049, People's Republic of China\\
% ...
$^{4}$ Carnegie Mellon University, Pittsburgh, Pennsylvania 15213, USA\\
% ...
$^{8}$ G.I. Budker Institute of Nuclear Physics SB RAS (BINP), Novosibirsk 630090, Russia\\
% ...
$^{19}$ (A)INFN Laboratori Nazionali di Frascati, I-00044, Frascati, Italy; (B)INFN and University of Perugia, I-06100, Perugia, Italy\\
% ...
$^{28}$ Peking University, Beijing 100871, People's Republic of China\\
% ...
$^{41}$ University of Rochester, Rochester, New York 14627, USA\\
$^{42}$ University of Science and Technology of China, Hefei 230026, People's Republic of China\\
% ...
$^{a}$ Also at the Novosibirsk State University, Novosibirsk, 630090, Russia\\
% ...
}
}

\date{\today}

\begin{abstract}

With data samples collected with the BESIII detector operating at
the BEPCII storage ring at center-of-mass energies from 4.009 to
4.420~GeV, the process $\EE\to \gamma\x$ is observed for the first time with a
statistical significance of $6.3\sigma$.  The measured mass of
the $\x$ is ($3871.9\pm 0.7_{\rm stat.}\pm 0.2_{\rm sys.}$)~MeV/$c^2$,
in agreement with previous measurements. Measurements of the product of the cross
section $\sigma[\EE\to \gamma\x]$ and the branching fraction
$\mathcal{B}[\x\to \ppjpsi]$ at center-of-mass energies 4.009,
4.229, 4.260, and 4.360~GeV are reported. Our
measurements are consistent with expectations for the radiative transition
process $\y\to \gamma\x$.

\end{abstract}


\maketitle

The $\x$ was first observed ten years ago by Belle~\cite{bellex} in $B^\pm\to
K^\pm\ppjpsi$ decays; it was subsequently confirmed
by several other experiments~\cite{CDFx,D0x,babarx}. Since its
discovery, the $\x$ has stimulated considerable interest. Both
{\em BABAR} and Belle observed the $\x\to \gamma\jpsi$ decay process,
which ensures that the $\x$ is a $C$-even
state~\cite{babar-jpc,belle-jpc}. The CDF and LHCb experiments
determined the spin-parity of the $\x$ to be
$J^{P}=1^{+}$~\cite{CDF-jpc,LHCbx},  and CDF also found that the
$\pp$ system was dominated by the $\rho^0(770)$
resonance~\cite{CDF-pp}.
Because of the proximity of its mass to the $\bar{D}D^*$ mass threshold,
the $\x$ has been interpreted as a candidate for a hadronic molecule or a tetraquark
state~\cite{epjc-review}. Until now, the $\x$ was only observed in $B$
meson decays and hadron collisions. Since the $\x$ is a $1^{++}$
state, it should be able to be produced through the radiative
transition of an excited vector charmonium or charmoniumlike
states such as a $\psi$ or a $Y$.

The puzzling $\y$~\cite{y4260} and $Y(4360)$~\cite{y4360}
vector charmoniumlike states have only been observed in
final states containing a charmonium meson and a $\pp$ pair, in contrast to
the $\psi(4040)$ and $\psi(4160)$ which dominantly couple to open charm final
states~\cite{pdg}.  The observation of the charged
charmoniumlike state $\z$~\cite{y4260,zc}, which is clearly not a
conventional charmonium state and is produced recoiling against
a $\pi^{\pm}$ at the CM energy of 4.26~GeV, indicates that these 
two ``exotic" states seem to couple with each other. To better understand their nature,
a investigation of other decay processes, such as the
radiative transition of the $\y$ and $Y(4360)$ to lower lying
charmonium or charmoniumlike states is important~\cite{model}.
The process $\y/Y(4360)\to \gamma\x$ is unique due to the exotic
feature of both the $\x$ and the $\y$ or $Y(4360)$ resonances.

In this Letter, we report the first  observation of the process $\EE\to \gamma\x\to
\gamma\ppjpsi$, $\jpsi \to \LL$ ($\LL=\EE$ or $\MM$) in an analysis
of data collected with the BESIII detector
operating at the BEPCII storage ring~\cite{bepc2} at $\EE$
center-of-mass (CM) energies from $\sqrt{s}=4.009$~GeV to
4.420~GeV~\cite{lum}. The CM energy is measured with a precision 
of $\pm 1.0$~MeV~\cite{BEMS}.
A {\sc geant4}-based Monte Carlo~(MC) simulation software package
that includes the geometric description of the BESIII detector and the
detector response is used to optimize the event selection
criteria, determine the detection efficiency, and estimate
backgrounds. For the signal process, we generate $\EE\to
\gamma\x$, with $\x\to \pp\jpsi$ at each CM energy. Initial
state radiation (ISR) is simulated with {\sc kkmc}~\cite{kkmc},
where the Born cross section of $\EE\to \gamma\x$ between 3.90 and
4.42~GeV is assumed to follow the $\EE\to \ppjpsi$
line-shape~\cite{y4260}. The maximum ISR photon energy corresponds
to the 3.9~GeV/$c^2$ production threshold of the $\gamma\x$
system. We generate $\x\to \rho^0\jpsi$ MC events with $\rho^0\to
\pp$ to model the $\pp$ system and determine the detection
efficiency~\cite{CDF-pp}. Here the $\rho^0$ and $\jpsi$ are assumed to
be in a relative $S$-wave. Final State Radiation (FSR) is handled with
{\sc photos}~\cite{photos}.



Events with four good charged tracks with net charge zero are
selected as described in Ref.~\cite{zc}.
%
Showers identified as photon candidates must satisfy fiducial and
shower quality as well as timing requirement as described in Ref.~\cite{eta-jpsi}.
When there is more than one photon candidate, the one with the largest
energy is regarded as the radiative photon.
In order to improve the momentum and energy resolution and reduce the background,
the event is subjected to a four-constraint~(4C) kinematic fit to the 
hypothesis $\EE\to \gamma\pi^+ \pi^- l^+ l^-$, that constrains
total four momentum of the measured particles to be equal to the initial
four-momentum of the colliding beams. The $\chi^2$ of the
kinematic fit is required to be less than 60.
To reject radiative Bhabha and radiative dimuon
($\gamma\EE/\gamma\MM$) backgrounds associated with
photon-conversion, the cosine of the opening angle of the pion
candidates, is required to be less than 0.98. This restriction
removes almost all the background events with an efficiency loss
for signal that is less than 1\%. Background from $\EE\to \eta\jpsi$
with $\eta\to \gamma\pp/\pp\pi^0$ is rejected by requiring
$M(\gamma\pp)>0.6~{\rm GeV}/c^2$, and its remaining contribution
is negligible~\cite{eta-jpsi, hjpsi-belle}.

After imposing the above requirements, there are clear $\jpsi$
peaks in the $\LL$ invariant mass distribution at each CM energy data set.
The $\jpsi$ mass window to select signal events is
$3.08<M(\LL)<3.12$~GeV/$c^2$ (mass resolution is 6~MeV/$c^2$),
while the sidebands are $3.0<M(\LL)<3.06$~GeV/$c^2$ and
$3.14<M(\LL)<3.20$~GeV/$c^2$, which is three times as wide
as the signal region.

\begin{figure}
\begin{center}
\includegraphics[height=3cm]{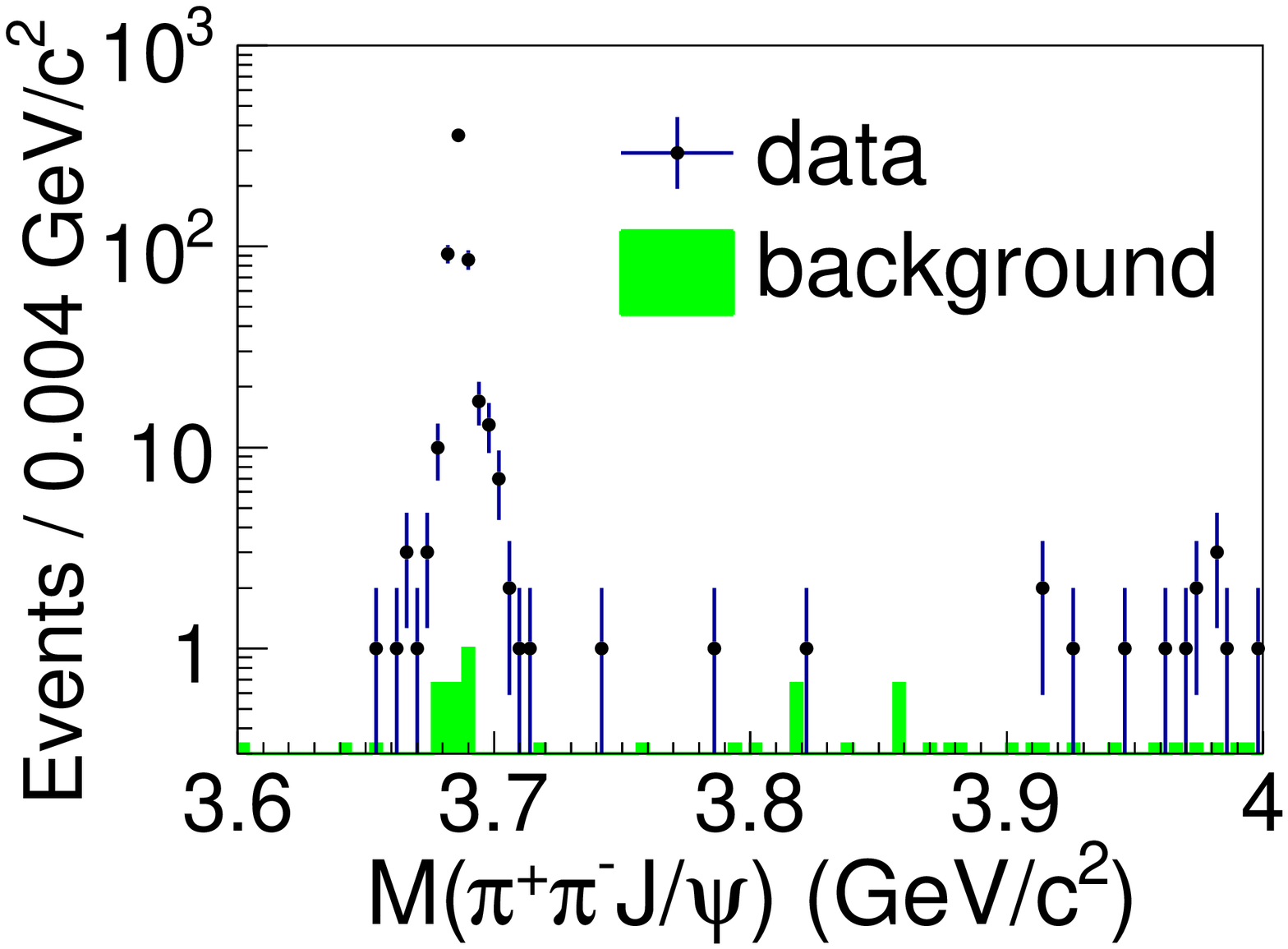}
\includegraphics[height=3cm]{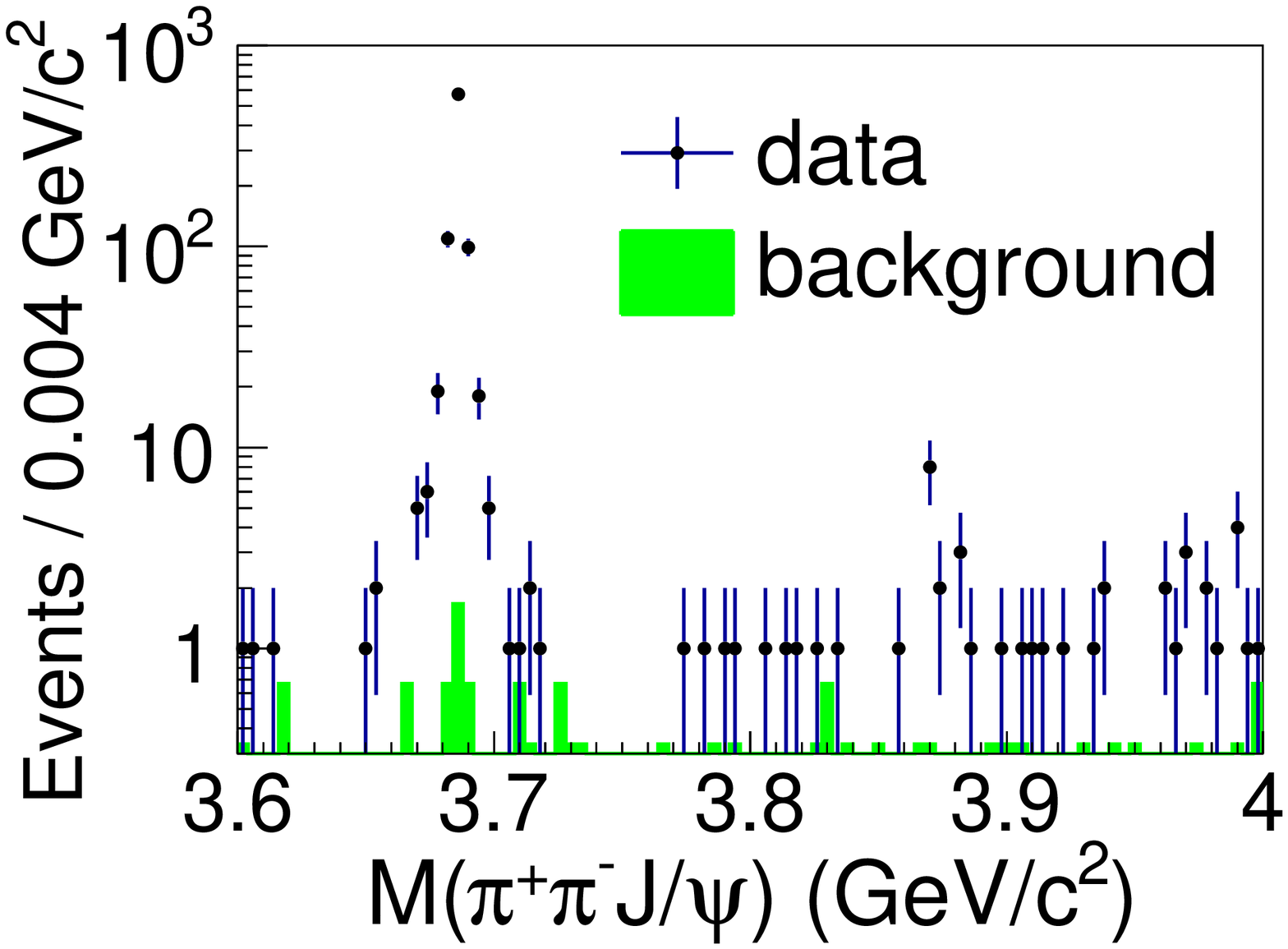}
\includegraphics[height=3cm]{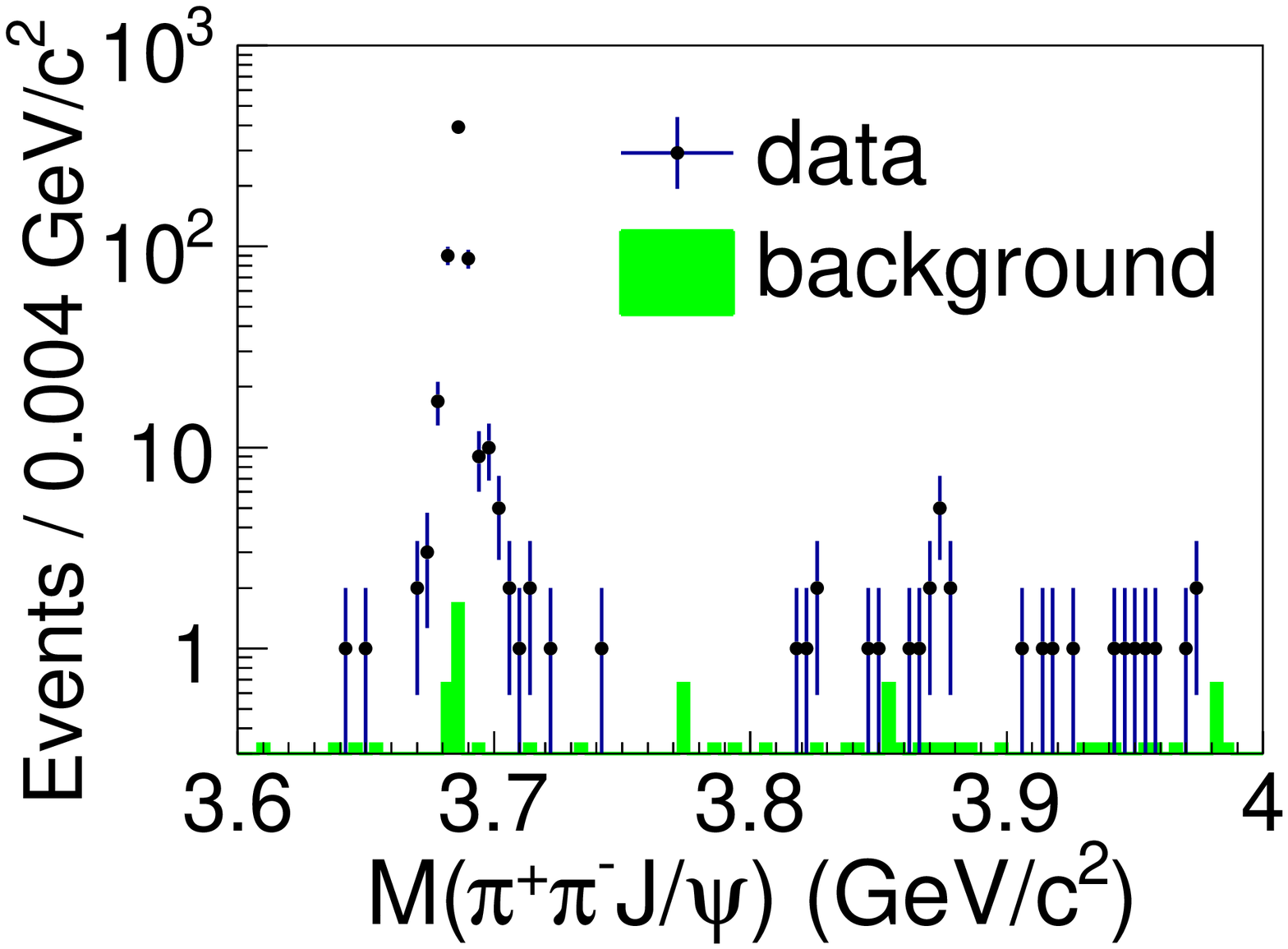}
\includegraphics[height=3cm]{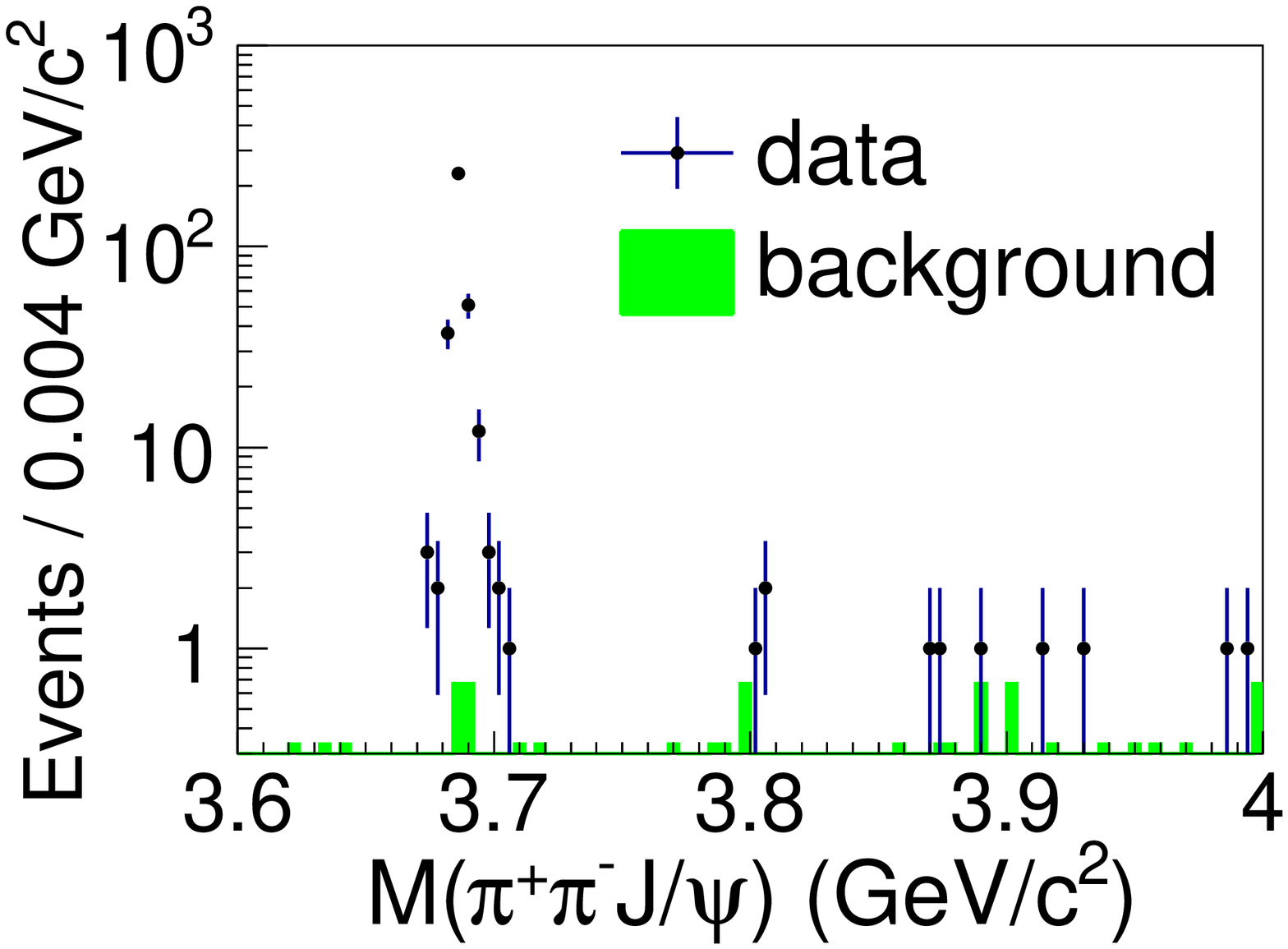}
\caption{The $\ppjpsi$ invariant mass distributions at $\sqrt{s}$ =
4.009 (top left), 4.229 (top right), 4.260 (bottom left), and
4.360~GeV (bottom right). Dots with error bars are data, the green
shaded histograms are normalized $\jpsi$ sideband events. }
\label{fig-mx3872}
\end{center}
\end{figure}

The remaining backgrounds mainly come from $\EE\to(\gamma_{\rm
ISR})\ppjpsi$, $\etap\jpsi$, and $\pp\pp\pi^0/\pp\pp\gamma$.
MC simulation based on available measurements for
$(\gamma_{\rm ISR})\ppjpsi$~\cite{y4260}, and cross sections measured from 
the same data samples for $\etap\jpsi$ ($\etap\to \gamma\pp/\pp\eta$)
shows a smooth, non-peaking $M(\ppjpsi)$ mass distribution in the $\x$ signal region,
and indicates that background from $\EE\to
\pp\pp(\pi^0/\gamma)$ is small and can be estimated
from the $\jpsi$ mass sideband data. Figure~\ref{fig-mx3872} shows the
$\ppjpsi$ invariant mass distributions at $\sqrt{s}=4.009$, 4.229,
4.260, and 4.360~GeV. Here $M(\ppjpsi)=M(\pp\LL)-M(\LL)+m(\jpsi)$
is used to reduce the resolution effect of the lepton pairs, and
$m(\jpsi)$ is the nominal mass of $\jpsi$~\cite{pdg}. There is a
huge $\EE\to \gamma_{\rm ISR}\psip$ signal at each CM energy data set.
In addition, there is a narrow peak around 3872~MeV/$c^2$ in the 4.229 and
4.260~GeV data samples, while there is no significant signal at
the other energies.

\begin{figure}
\begin{center}
\includegraphics[height=6cm]{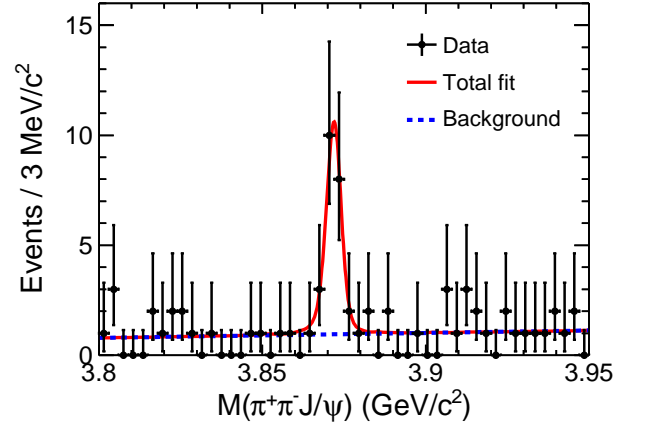}
\caption{Fit of the $M(\ppjpsi)$ distribution with a MC simulated
histogram convolved with a Gaussian function for signal and a
linear background function. Dots with error bars are data, the red
curve shows the total fit result, while the blue dashed curve
shows the background contribution.} \label{fit-mx}
\end{center}
\end{figure}

The $M(\ppjpsi)$ distribution (summed over all CM energy data sets) is
fitted to determine the mass and $\x$ yield. We use a MC simulated
signal histogram convolved with a Gaussian function which
represents the resolution difference between data and MC
simulation as the signal shape, and a linear function for the
background. The ISR $\psip$ signal is used to calibrate the
absolute mass scale and to extract the resolution difference
between data and MC simulation.  The fit to the $\psip$ results in
a mass shift of $\mu_{\psip}=-(0.34\pm 0.04)$~MeV/$c^2$, and a
standard deviation of the  Gaussian resolution function of
$\sigma=(1.14\pm 0.07)$~MeV/$c^2$.  The resolution parameter of
the resolution Gaussian applied to the MC simulated signal shape
is fixed at 1.14~MeV/$c^2$ in the fit to the $\x$.
Figure~\ref{fit-mx} shows the fit result (with $M[\x]_{\rm
input}=3871.7$~MeV/$c^2$ as input in MC simulation), which gives
$\mu_{\x}=-(0.10\pm 0.69)$~MeV/$c^2$ and $N[\x]=20.1\pm 4.5$. So,
the measured mass of $\x$ is $M[\x] = M[\x]_{\rm input} + \mu_{\x}
- \mu_{\psip} = (3871.9\pm 0.7)$~MeV/$c^2$, where the uncertainty
includes the statistical uncertainties from the fit and the mass
calibration. The limited statistics prevent us from
measuring the intrinsic width of the $\x$. From a fit with a
floating width we obtain $\Gamma[\x]=(0.0^{+1.7}_{-0.0})$~MeV, or less
than 2.4~MeV at the 90\% confidence level (C.L.). The statistical
significance of $\x$ is $6.3\sigma$, estimated by comparing the
difference of log-likelihood value [$\Delta(-2\ln\mathcal{L}) =
44.5$] with and without the $\x$ signal in the fit, and taking the
change of the number-of-degrees-of-freedom ($\Delta$ndf=2) into
consideration.

\begin{figure}
\begin{center}
\includegraphics[height=3.0cm]{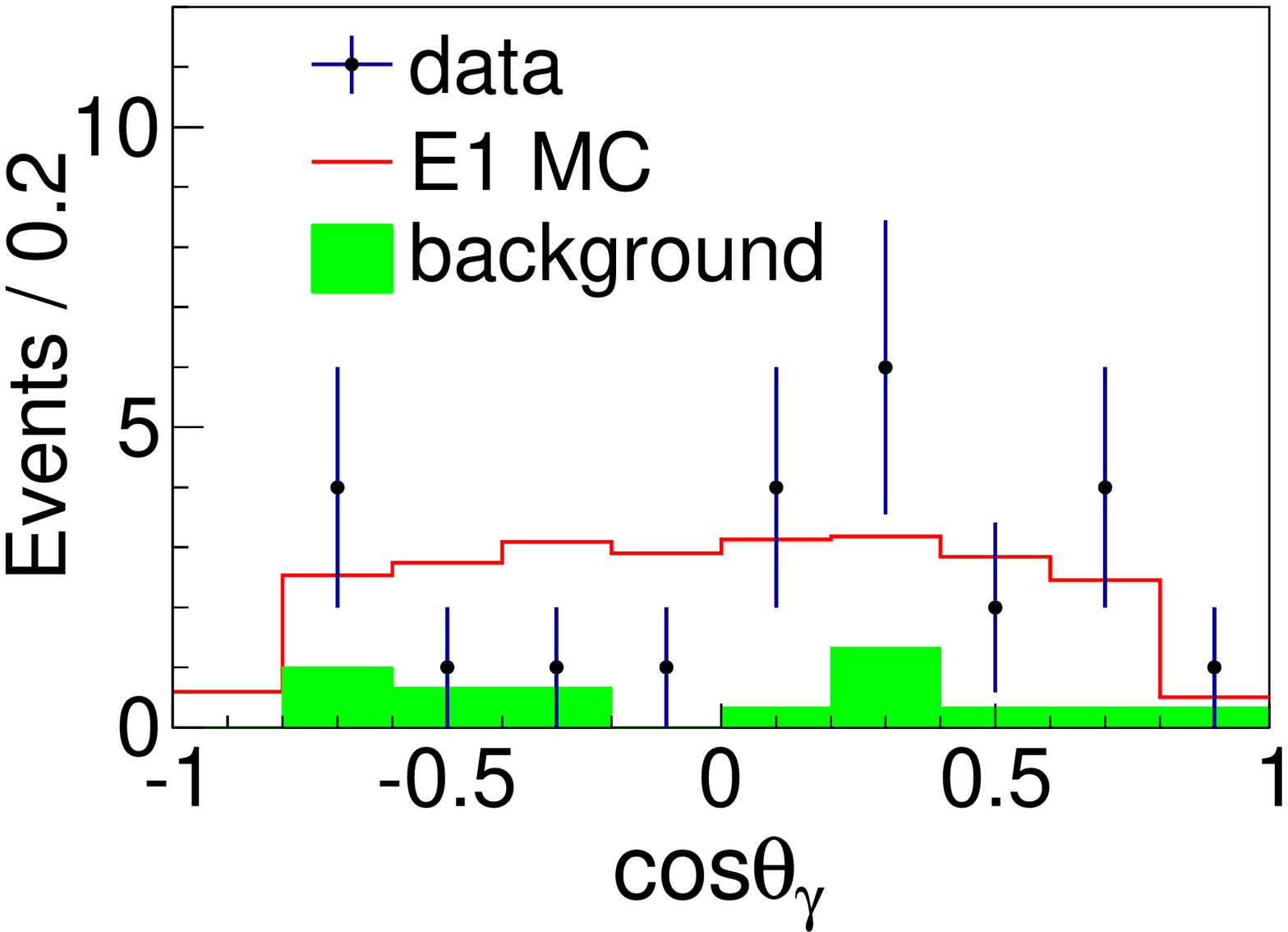}
\includegraphics[height=3.0cm]{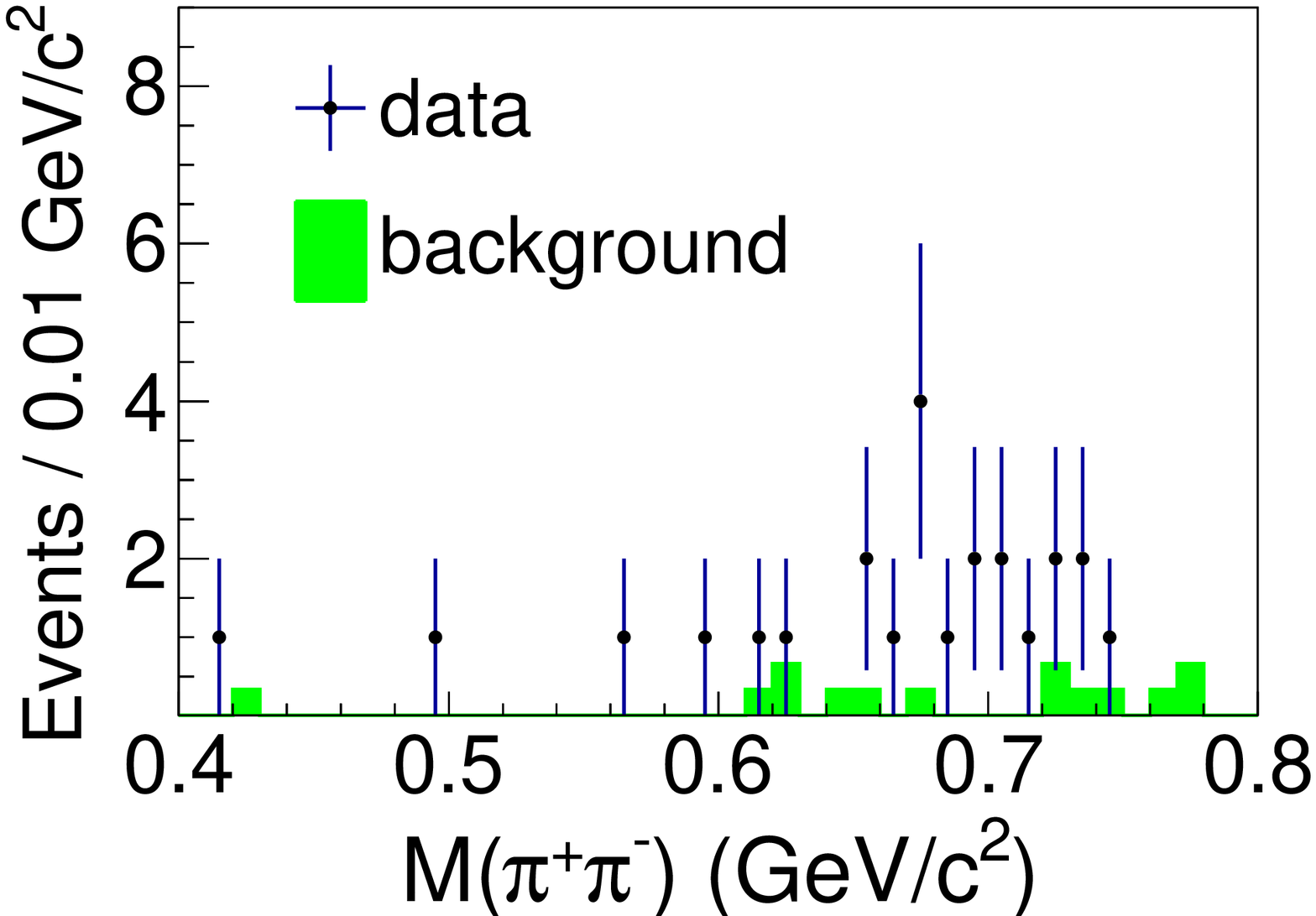}
\caption{The $\cos\theta$ distribution of the radiative
photon in $\EE$ CM frame (left) and the $M(\pp)$ distribution (right).
Dots with error bars are data in the $\x$ signal region, the green
shaded histograms are normalized $\x$ sideband events, and the red
open histogram in the left panel is the result from a MC simulation
that assumes a pure $E1$-transition.}  \label{cos-gam}
\end{center}
\end{figure}

Figure~\ref{cos-gam} shows the angular distribution
of the radiative photon in the $\EE$ CM frame and the $\pp$
invariant mass distribution, for the $\x$ signal events ($3.86 <
M(\ppjpsi) < 3.88$~GeV/$c^2$) and normalized sideband events
($3.83 < M(\ppjpsi) < 3.86$~GeV/$c^2$ or $3.88 < M(\ppjpsi) <
3.91$~GeV/$c^2$). The data agree with MC simulation assuming a pure
$E1$-transition between the $\y$ and the $\x$ for the polar angle
distribution, and the $M(\pp)$ distribution is consistent with
the CDF observation~\cite{CDF-pp} of a dominant $\rho^0(770)$
resonance contribution.

\begin{figure}
\begin{center}
\includegraphics[height=6cm]{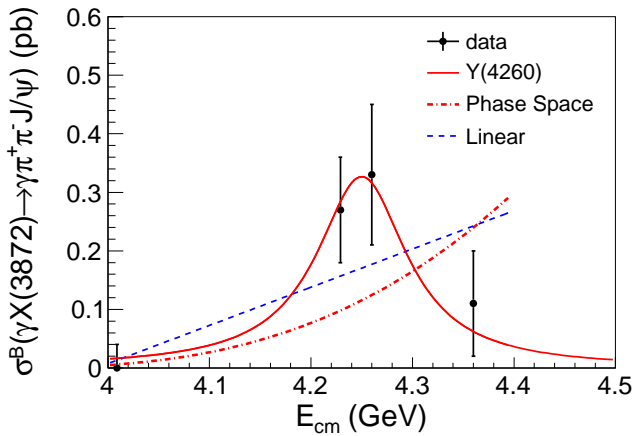}
\caption{The fit to $\sigma^B[\EE\to \gamma\x]\times
\mathcal{B}[\x\to \ppjpsi]$ with a $\y$ resonance (red solid
curve), a linear continuum (blue dashed curve), or a
$E1$-transition phase space term (red dotted-dashed curve). Dots
with error bars are data.} \label{xsec}
\end{center}
\end{figure}

The product of the Born-order cross section times the branching
fraction of $\x\to\ppjpsi$ is calculated using
$\sigma^{B}[\EE\to\gamma \x]\times \mathcal{B}[\x\to\ppjpsi] =
\frac{N^{\rm obs}} {\mathcal{L}_{\rm int} (1+\delta) \epsilon
\mathcal{B}}$, where $N^{\rm obs}$ is the number of observed
events obtained from the fit to the $M(\ppjpsi)$ distribution,
$\mathcal{L}_{\rm int}$ is integrated luminosity, $\epsilon$ is
the detection efficiency, $\mathcal{B}$ is the branching fraction
of $\jpsi\to \LL$ and ($1+\delta$) is the radiative correction
factor, which depends on the line shape of $\EE\to
\gamma\x$. Since we observe large cross sections at
$\sqrt{s}=4.229$ and 4.260~GeV, we assume the $\EE\to \gamma\x$
cross section follows that of $\EE\to \ppjpsi$ over the full energy range of
interest and use the $\EE\to \ppjpsi$ line-shape from published
results~\cite{y4260} as input in the calculation of the efficiency
and radiative correction factor. The results of these studies at
different energies
($\sqrt{s}=4.009$~GeV, 4.229~GeV, 4.260~GeV, and 4.360~GeV) are
listed in Table~\ref{sec}. For the 4.009~GeV and 4.360~GeV data, where
the $\x$ signal is not statistically significant, upper limits for production
yield at 90\% C.L. are also given. As a validation, the measured
ISR $\psip$ cross section at each energy, together with the
corresponding QED prediction~\cite{isr-psip} are also listed in
Table~\ref{sec}, where there is good agreement.

\begin{table*}
\begin{center}
\caption{The number of $\x$ events ($N^{\rm obs}$),
radiative correction factor ($1+\delta$), detection efficiency
($\epsilon$), measured Born cross section $\sigma^{B}[\EE\to
\gamma\x]$ times $\mathcal{B}[\x\to \ppjpsi]$ ($\sigma^B\cdot
{\cal B}$, where the first uncertainties are statistical and the second
systematic), measured ISR $\psip$ cross section ($\sigma^{\rm
ISR}$, where the first uncertainties are statistical and the second
systematic), and predicted ISR $\psip$
cross section ($\sigma^{\rm QED}$ with uncertainties from resonant
parameters) from QED~\cite{isr-psip} using resonant parameters in
PDG~\cite{pdg} as input at different energies. For 4.009~GeV
and 4.360~GeV, the upper limits of observed events ($N^{\rm up}$)
and cross section times branching fraction ($\sigma^{\rm
up}\cdot\mathcal{B}$) are given at the 90\% C.L.} \label{sec}
\begin{tabular}{ccccccccc}
  \hline\hline
  $\sqrt{s}$~(GeV) & $N^{\rm obs}$ & $N^{\rm up}$ & $\epsilon$ (\%) & $1+\delta$
  & $\sigma^B\cdot\mathcal{B}$~(pb) & $\sigma^{\rm up}\cdot\mathcal{B}$~(pb) & $\sigma^{\rm ISR}$~(pb) & $\sigma^{\rm QED}$~(pb) \\
  \hline
  4.009 & $0.0\pm 0.5$ & $<1.4$ & 28.7 & 0.861 & $0.00\pm0.04\pm0.01$ & $<0.11$ & $719\pm 30 \pm47$ & $735\pm13$ \\
  4.229 & $9.6\pm 3.1$ &  -          & 34.4 & 0.799 & $0.27\pm0.09\pm0.02$ & -             & $404\pm 14 \pm27$ & $408\pm7$ \\
  4.260 & $8.7\pm 3.0$ &  -          & 33.1 & 0.814 & $0.33\pm0.12\pm0.02$ & -             & $378\pm 16 \pm25$ & $382\pm7$ \\
  4.360 & $1.7\pm 1.4$ & $<5.1$ & 23.2 & 1.023 & $0.11\pm0.09\pm0.01$ & $<0.36$ & $308\pm 17 \pm20$ & $316\pm5$ \\
  \hline\hline
\end{tabular}
\end{center}
\end{table*}

We  fit the energy-dependent cross section with
a $\y$ resonance (parameters fixed to PDG~\cite{pdg} values), a
linear continuum, or a $E1$-transition phase space ($\propto
E^3_\gamma$) term. Figure~\ref{xsec} shows all the fit results,
which give $\chi^2/{\rm ndf}=0.49/3$ (C.L.=92\%), 5.5/2
(C.L.=6\%), and 8.7/3 (C.L.=3\%) for  a $\y$ resonance, linear
continuum, and phase space distribution, respectively.
The $\y$ resonance describes the data better than the
other two options.


The systematic uncertainty in the $\x$ mass measurement include those
from the absolute mass scale and the parametrization of the $\x$
signal and background shapes. Since we use ISR $\psip$ events to
calibrate the fit, the systematic uncertainty from the mass scale
is estimated to be 0.1~MeV/$c^2$ (including statistical uncertainties of
the MC samples used in calibration procedure). In the $\x$ mass
fit, a MC simulated histogram with a zero width is used to
parameterize the signal shape. We replace this histogram with a
simulated $\x$ resonance with a width of $1.2$~MeV~\cite{pdg} (the
upper limit of the $\x$ width at 90\% C.L.) and repeat the fit;
the change in mass for this new fit is taken as the systematic uncertainty due to the
signal parametrization, which is 0.1~MeV/$c^2$. Likewise, changes measured
with a background shape from MC-simulated $(\gamma_{\rm ISR})\ppjpsi$ and
$\etap\jpsi$ events indicate a systematic uncertainty associated with the background
shape of  0.1~MeV/$c^2$ in mass. By summing the contributions from all sources assuming
that they are independent, we obtain a total systematic uncertainty of
0.2~MeV/$c^2$ for the $\x$ mass measurement.

The systematic uncertainty in the cross section measurement mainly comes
from efficiencies, signal parametrization, background shape,
radiative correction, and luminosity measurement. The luminosity
is measured using Bhabha events, with an uncertainty of 1.0\%. The
uncertainty of tracking efficiency for high momenta leptons is
1.0\% per track. Pions have momentum ranges from 0.1 to
0.6~GeV/$c$ at $\sqrt{s}=4.260$~GeV, and with a small change with
different CM energies. The momentum-weighted uncertainty is also estimated to be
1.0\% per track. In this analysis, the radiative
photons have energies that several hundreds of MeV. Studies with a
sample of $\jpsi\to \rho\pi$ events show that the uncertainty in the reconstruction efficiency
for photons in this energy range is less than 1.0\%.

The number of $\x$ signal events is obtained through a fit to the
$M(\ppjpsi)$ distribution. In the nominal fit, a simulated
histogram with zero width convolved with a Gaussian function is
used to parameterize the $\x$ signal. When a MC-simulated signal
shape with $\Gamma[\x]=1.2$~MeV~\cite{pdg} is used, the difference
in the $\x$ signal yield, is 4.0\%; this is taken as the systematic
uncertainty due to signal parametrization.
Changing the background shape from a linear term to the expected
shape from the dominant background source $\etap\jpsi$ results in
a 0.2\% difference in the $\x$ yields. The $\EE\to \ppjpsi$ line
shape affects the radiative correction factor and detection
efficiency. Using the measurements from BESIII, Belle, and {\em
BABAR}~\cite{y4260} as inputs, the maximum difference in
$(1+\delta)\epsilon$ is 0.6\%, which is taken as the systematic
uncertainty. The uncertainty from the kinematic fit is estimated with
the very pure ISR $\psip$ sample, and the efficiency difference between
data and MC simulation is found to be 1.5\%.
The systematic uncertainty for the $\jpsi$ mass window is also estimated
using the ISR $\psip$ events, and
the efficiency difference between data and MC simulation is found
to be ($0.8\pm0.8$)\%. We conservatively take 1.6\% as the systematic uncertainty
due to $J/\psi$ mass window. The
uncertainty in the branching fraction of $\jpsi\to \LL$ is taken
from Ref.~\cite{pdg}. The efficiencies for other selection
criteria, the trigger simulation, the event start time
determination, and the final-state-radiation simulation are quite
high ($>99\%$), and their systematic uncertainties are estimated to be
less than 1\%.
Assuming all the systematic uncertainty sources are independent, we add
all of them in quadrature, and the total systematic uncertainty is
estimated to be 6.5\%.

In summary, we report the first observation of the
process $\EE\to \gamma\x$. The
measured mass of the $\x$, $M[\x]=(3871.9\pm 0.7\pm 0.2)~{\rm
MeV}/c^2$, agrees well with previous measurements~\cite{pdg}. The
production rate $\sigma^B[\EE\to \gamma\x]\cdot \mathcal{B}[\x\to
\ppjpsi]$ is measured to be ($0.27\pm 0.09\pm 0.02$)~pb at
$\sqrt{s}=4.229$~GeV, ($0.33\pm 0.12\pm 0.02$)~pb at
$\sqrt{s}=4.260$~GeV, less than $0.11$~pb at $\sqrt{s}=4.009$~GeV,
and less than $0.36$~pb at $\sqrt{s}=4.360$~GeV at the 90\% C.L.
Here the first uncertainties are statistical and the second systematic.
(For the upper limits, the efficiency has been lowered by a factor
of ($1-\sigma_{\rm sys}$).)

These observations strongly support the existence of
the radiative transition process $\y\to \gamma\x$. 
While the measured cross sections at around 4.260~GeV are an order 
of magnitude higher than the NRQCD calculation of continuum production~\cite{ktchao},
the resonant contribution with $\y$ line shape provides a better description
of the data than either a linear continuum or a $E1$-transition
phase space distribution. The $\y\to \gamma\x$ could be another
previously unseen decay mode of the $\y$ resonance. This, together
with the previously reported transitions to the charged
charmoniumlike state $\z$ (which is manifestly exotic)~\cite{y4260,zc},
suggest that there might be some commonality in the nature of these
three different states. This may be a clue that can facilitate
a better theoretical interpretation of them. As an example, the
measured relative large $\gamma\x$ production rate near 4.260~GeV
is similar to the model dependent calculations in Ref.~\cite{model}
where the $\y$ is taken as a $\bar{D}D_1$ molecule.

Combining with the $\EE\to \ppjpsi$ cross section
measurement at $\sqrt{s}=4.260$~GeV from BESIII~\cite{zc}, we
obtain $\sigma^B[\EE\to \gamma\x]\cdot \mathcal{B}[\x\to
\ppjpsi]/\sigma^B(\EE\to \ppjpsi) = (5.2\pm 1.9)\times 10^{-3}$,
under the assumption that the $\x$ is produced only from the $\y$
radiative decays and the $\pp\jpsi$ is only from the $\y$ hadronic
decays. If we take $\mathcal{B}[\x\to \ppjpsi]=5\%$~\cite{bnote},
then $\mathcal{R} = \frac{\sigma^B[\EE\to \gamma\x]}
{\sigma^B(\EE\to \ppjpsi)}=0.1$, or equivalently,
$\frac{\mathcal{B}[\y\to \gamma\x]} {\mathcal{B}(\y\to
\ppjpsi)}=0.1$.

The BESIII collaboration thanks the staff of BEPCII and the computing center for their strong support. This work is supported in part by the Ministry of Science and Technology of China under Contract No. 2009CB825200; National Natural Science Foundation of China (NSFC) under Contracts Nos. 10625524, 10821063, 10825524, 10835001, 10935007, 11125525, 11235011; Joint Funds of the National Natural Science Foundation of China under Contracts Nos. 11079008, 11179007; the Chinese Academy of Sciences (CAS) Large-Scale Scientific Facility Program; CAS under Contracts Nos. KJCX2-YW-N29, KJCX2-YW-N45; 100 Talents Program of CAS; German Research Foundation DFG under Contract No. Collaborative Research Center CRC-1044; Seventh Framework Programme of the European Union under Marie Curie International Incoming Fellowship Grant Agreement No. 627240; Istituto Nazionale di Fisica Nucleare, Italy; Ministry of Development of Turkey under Contract No. DPT2006K-120470; U. S. Department of Energy under Contracts Nos. DE-FG02-04ER41291, DE-FG02-05ER41374, DE-FG02-94ER40823, DESC0010118; U.S. National Science Foundation; University of Groningen (RuG) and the Helmholtzzentrum fuer Schwerionenforschung GmbH (GSI), Darmstadt; WCU Program of National Research Foundation of Korea under Contract No. R32-2008-000-10155-0.


\end{document}